\documentclass[journal]{IEEEtran}
\usepackage{float}
\usepackage{here}

\usepackage{cite}
\usepackage[linesnumbered,ruled,vlined]{algorithm2e}
\usepackage{amsmath}
\usepackage{amsthm}
\usepackage[ruled,vlined]{algorithm2e}
\theoremstyle{remark}

\newtheorem*{remark*}{Remark}

\newtheorem{theorem}{Theorem}
\usepackage{algpseudocode}

\ifCLASSINFOpdf
  \usepackage[pdftex]{graphicx}
\else
   \usepackage[dvips]{graphicx}
\fi
\ifCLASSOPTIONcompsoc
 \usepackage[caption=false,font=normalsize,labelfont=sf,textfont=sf]{subfig}
\else
\usepackage[caption=false,font=footnotesize]{subfig}
\begin{document}
%
\title{Analysis of Distributed Average Consensus Algorithms for Robust IoT networks}
%
%
%

\author{Sateeshkrishna Dhuli and
        {Fouzul Atik}
}

%
%

\markboth{}%
{Shell \MakeLowercase{\textit{et al.}}: }
%



\maketitle


\begin{abstract}
Internet of Things(IoT) is a heterogeneous network consists of various physical objects such as large number of sensors, actuators, RFID tags, smart devices, and servers connected to the internet. IoT networks have potential applications in healthcare, transportation, smart home, and automotive industries. To realize the IoT applications, all these devices need to be dynamically cooperated and utilize their resources effectively in a distributed fashion. Consensus algorithms have attracted much research attention in recent years due to their simple execution, robustness to topology changes, and distributed philosophy. These algorithms are extensively utilized for synchronization, resource allocation, and security in IoT networks. Performance of the distributed consensus algorithms can be effectively quantified by the Convergence Time, Network Coherence, Maximum Communication Time-Delay.  In this work, we model the IoT network as a q-triangular r-regular ring network as q-triangular topologies exhibit both small-world and scale-free features. Scale-free and small-world topologies widely applied for modelling IoT as these topologies are effectively resilient to random attacks. In this paper, we derive explicit expressions for all eigenvalues of Laplacian matrix for q-triangular r-regular networks. We then apply the obtained eigenvalues to determine the convergence time, network coherence, and maximum communication time-delay. Our analytical results indicate that the effects of noise and communication delay on the consensus process are negligible for q-triangular r-regular networks. We argue that q-triangulation operation is responsible for the strong robustness with respect to noise and communication time-delay in the proposed network topologies.

\end{abstract}

\IEEEpeerreviewmaketitle

\section{Introduction}
\IEEEPARstart{I}{nternet} of Things is formed by a large number of smart devices that dynamically cooperate and make their resources available to achieve a common goal. Smart devices consist of large number of sensors, actuators, RFID tags, smart devices, and servers connected to the internet. These devices should be able to take decisions, cooperative, and exchange information with the other devices and human beings. Consensus algorithms are simple distributed algorithms can be employed for the synchronization, security, and resource allocation problems in IoT Networks. These algorithms are suitable to resource constrained, dynamic, and fault tolerant networks. These algorithms performance is measured by the Convergence Time, Network Coherence, and Maximum Communication Time-Delay \cite{saber2003consensus}, \cite{xiao2007distributed}. Convergence Time is defined as the time required by the nodes to converge to the average of the initial node values. Convergence time is evaluated by the second smallest eigenvalue of Laplacian matrix. In consensus algorithms, in the absence of noise, nodes converge asymptotically to average of the initial state values. Due to noise, the node values fluctuates and they will not converge to the average, which can be studied by the network coherence.

Network coherence \cite{patterson2014consensus}, \cite{bamieh2012coherence} measures the deviation of node values from average of the initial state values. It is effectively captures the variance of the fluctuations in the first-order consensus system. In the second-order system, each node has two state variables, where the state of the entire system is thus captured by two vectors. Communication Time-Delay \cite{olfati2004consensus} measures ability of consensus algorithm resistant to communication delay between nodes, which is decided by the largest eigenvalue of Laplacian matrix. All these metrics depend on the eigenvalues of the Laplacian matrix, which represents the topology of any network. To avail the advantages of consensus algorithms, it is necessary to measure these performance metrics for IoT networks. To understand the effect of network parameters on convergence time of consensus algorithm, authors modeled the WSN as an r-nearest neighbor network in \cite{dhuli2015convergence}. This work shown that convergence time can be increased drastically by increasing the number of nearest neighbors. Gossip algorithm is a special case of consensus algorithm. In \cite{kouachi2020convergence}, authors modeled the WSN as a one dimensional lattice network and derive the explicit expressions of convergence rate for periodic gossip algorithms. Lattice networks model the finite sized resource constrained networks and facilitate the closed form expressions for convergence time. 

Small-world and Scale-free topologies are popular complex network models that can be applied to IoT networks to improve the robustness against random networks. Small-world model creates shorter average path length which reduces the maximum communication delay drastically in IoT. Moreover, adding shortcuts in communication will also improves the robustness. Scale-free model \cite{chen2019intelligent} is one of the classic models in which the node degree follows power-law distribution. These topologies give better performance in withstanding random attacks. Because of these salient features, small-world and scale-free topologies have been utilized in modeling IoT networks \cite{qiu2018tosg},  \cite{qiu2017data}, \cite{sohn2017small}. Triangulation is a popular graph operation \cite{yi2018scale}, \cite{zeng2018hitting} in network science to obtain network models for studying the scale-free and small-world characteristics of networks. In our work, we modeled the IoT network as a q-triangular r-regular ring network and derived the closed-form expressions of convergence time, network coherence, and maximum communication time-delay. Our numerical results depict that our proposed network topology significantly improves the robustness of the consensus algorithms against communication delays and noise. 
\section{Related Work}
In literature, consensus algorithms have been widely studied for IoT networks. In particular, most of the researchers used these algorithms for distributed computation, security, and synchronization problems. In \cite{carvin2014generalized}, authors employed the consensus algorithm for distributed monitoring of IoT networks. This work used f-consensus theory and derived convergence conditions that allows a nice trade-off between precision and resource consumption. A consensus based distributed algorithm for service detection and data processing is proposed in \cite{li2014distributed} for IoT networks. This algorithm calculate the consensus locally and combines in an iterative fashion to improves the robustness of the consensus process. Resource allocation, task allocations are critical issues to be focused in resource constrained networks. In \cite{colistra2014problem}, authors propose a consensus based optimization algorithm for resource allocation in heterogeneous IoT networks. They have shown that network converges to a solution where network resources are homogeneously distributed.

A distributed algorithm where objects cooperate to reach a consensus on resources allocation is proposed in \cite{pilloni2017consensus}. This work implements optimization process to select the objects that would guarantee the minimum Quality of Information and improve the lifetime of objects. Authors in \cite{orostica2018robust}, evaluated the average consensus algorithm on IoT testbed. This work avoids the deadlock problems and deals with packet losses, delays, and multi-rate behavior in IoT networks. A distributed soft clustering algorithm based on average consensus algorithm for the IoT is presented in \cite{yu2020distributed}. This work claimed it provides the stable clustering quality for IoT networks. Blockchain technology is well known for its potential use in security mechanisms and protect from different attacks. However, this technology is suitable to power constrained devices. Consensus algorithms have been extensively used in blockchains for IoT networks. In \cite{salimitari2020survey}, authors presented the survey of the various blockchain based consensus methods that are suitable to resource constrained IoT networks. 

In \cite{huang2019towards}, authors proposed a blockchain system with credit-based consensus mechanism for IoT networks which ensures system security and transaction efficiency simultaneously. Trust models are very popular in ensuring security for peer to peer, WSN, and IoT networks. In \cite{ma2020towards}, author proposes a trust model based on consensus to evaluate the trustworthiness of IoT nodes to detect malicious nodes. Authors propose a novel consensus algorithm called Proof-of-Authentication in \cite{puthal2020poah} to introduce a cryptographic authentication mechanism for resource constrained devices. Authors proposed a lightweight proof of block and trade consensus algorithm for IoT blockchain in \cite{biswas2019pobt}. In \cite{chen2020consensus}, authors presented a distributed clustering algorithm for IoT networks where observations are distributed and data transmission is only allowed between one-hop neighbors. A  blockchain based on consensus algorithm for IoT applications is proposed in \cite{dorri2020tree}. In \cite{salimitari2018survey}, authors presented a survey on blockchain-based consensus methods for resource-constrained IoT networks. Authors presented a lightweight consensus algorithm in \cite{maitra2020integration} that can be implemented in the IoT environment. This algorithm evaluates the feasibility of medical supply and drug transportation to mitigate privacy issues. In our work, we provide the theoretical tools to study the consensus algorithms for IoT networks. The properties of small-world and scale-free networks ensure the network robustness against random attacks. This motivates us to develop the q-triangular r-regular network which incorporates the properties of both small-world and scale-free networks. We derive the closed form expressions of convergence time, network coherence, and maximum communication-time delay and study the effect of network parameters on these performance metrics. Our theoretical results provide important insights to design and control the convergence of consensus algorithms.
 \subsection{Main Contributions}
1) Firstly, we model the IoT network as a q-triangular r-regular ring network and compute the Laplacian eigenvalues. \\
2) Secondly, we derive the explicit expressions for convergence time, first order network coherence, second order network coherence, and maximum communication time-delay of q-triangular r-regular ring networks for average consensus algorithms.\\
3) Finally, we present the numerical results and study the effect of node degree, network size, triangulation parameter on convergence time, network coherence, and maximum communication time-delay of consensus algorithms. 
\subsection{Organization}
The remainder of the paper is organized as follows. We provide some preliminaries and review the consensus algorithms in Section III. In Section IV, we derive the explicit eigenvalues of consensus algorithm for q-triangular r-regular networks. In Section V, we derive the explicit expressions of convergence time, network coherence, and maximum communication time-delay. We present the numerical results and study the effect of triangulation parameter, network size, and node degree on the convergence time, network coherence, and maximum communication time-delay of consensus algorithms in Section VI. Finally, we discuss the conclusions in Section VII. \\

\textit{Notations}: Table 1 presents the notations and corresponding definitions used in the paper.
\begin{table}
\centering
\caption{List of Notations}
\begin{tabular}{ |l|l| } 
\hline
Notation & Definition\\
  \hline
  $\textbf{x}$& Vector of a state variables\\
$n$&Number of nodes \\
$\lambda_2$&Second Smallest Eigenvalue \\
$W$& Weight matrix \\
$T$&Convergence Time \\
$\lambda$&Eigen Value \\
$h$&Consensus parameter \\
$\gamma$&Convergence parameter \\
$B$&Incidence Matrix \\
$H^{(1)}$&First Order Network Coherence\\ 
$H^{(2)}$&Second Order Network Coherence\\ 
$T_{\max }$&Maximum Communication Time-Delay\\
$r$&Node Degree\\
$q$&Triangulation Parameter \\
$h$&Consensus Parameter \\
$\lambda _{n-1}$& Largest Eigenvalue\\
\hline
\end{tabular}
\end{table}
\section{Brief Review of Average Consensus Algorithm}
In this section, we introduce some basic concepts in spectral graph
theory and review the consensus algorithm. We consider an $n$-vertex simple connected graph $G=(V,E)$, where$V=V(G)=\{1,2,\dots,n\}$ is the vertex set and
$E=E(G)=\{e_1,e_2,\dots,e_m\}$ is the edge set of the graph. The
\emph{adjacency matrix} $A(G)$ of $G$ is a square matrix of order
$n$, whose $(i,j)^{th}$ entry is equal to 1(or 0) if the vertices
$i$ and $j$ are adjacent(or not adjacent). The \emph{incidence
matrix} $B$ of $G$ is a matrix of order $n\times m$, whose
$(i,j)^{th}$ entry is equal to 1(or 0) if the vertex $i$ and the
edge $e_j$ are incident(or not incident). Let $D(G)$ be the diagonal
matrix of vertex degree. Then the \emph{Laplacian matrix} of $G$ is
$L(G)=D(G)-A(G)$. Also we have $L(G)=2D(G)-BB^T$. \\

Let $x_k (0)$ denotes the real scalar variable of node $k$ at $t=0$. Average consensus algorithm computes the average $x_{avg}=\frac{{\sum\nolimits_{k = 1}^n {x_k (0)} }}{n}$ at every node through a distributed approach which does not require any centralized node. At time instant $t+1$, the real scalar variable at node `\textit{i}' is expressed as 
\begin{equation}
x_k (t + 1) = x_k (t) + h\sum\limits_{j \in N_k } {(x_j (t) - x_k (t))} ,\,\,\,k = 1,...,n,
\label{2}
\end{equation}
where `\textit{h}' is a consensus parameter and $N_k$ denotes neighbor set of node `\textit{k}'.
This can be also expressed as a simple linear iteration as 
\begin{equation}
\textbf{x(t + 1)} = W\textbf{x(t)},\,\,\,\,t = 0,1,2...,
\label{3}
\end{equation}
where `\textit{W}' denotes weight matrix, and $W_{kj}$ is a weight associated with the edge $(k,j)$. If we assign equal weight `\textit{h}' to each link in the network, then optimal weight for a given topology is
\begin{equation}
W_{kj}=\left\{\begin{matrix}
h & if ,\,\,\,\,(k,j) \in E,\\
1-hdeg(\nu_{k})&if,\,\,\,\,k=j,\\
0 & otherwise.
\end{matrix}\right.
\label{4}
\end{equation}
and weight matrix is given by
\begin{equation}
W = I - hL.
\label{5}
\end{equation}
where `\textit{I}' is a $n\times n$ identity matrix. \\

Definition 1: Convergence time ($T$) \cite{xiao2007distributed} is defined as the time required for nodes to reach the consensus. And it is  measured by the 
\begin{equation}
T = \frac{1}{{\ln \left( {\frac{1}{\gamma }} \right)}}, \label{e11}
\end{equation}
where $\gamma$ is convergence parameter. \\

Physical objects in IoT network should be robust with respect to different parameters, including hardware failure, environmental uncertainty and communication failures. Robustness to uncertainty and noise can be effectively measured by network coherence \cite{young2010robustness}, \cite{qi2018consensus}. \\

Definition 2: Network coherence is also defined as robustness to noise, and it can be measured by the deviation of each node’s state from the global average of all current states. \\
In the first-order consensus problem, each node has a single state subject to noise.
Network coherence of a first order system is measured by
\begin{equation}
H^{(1)}  = \frac{1}{{2N}}\sum\limits_{k = 2}^N {\frac{1}{{f(\lambda_k) }}} \label{e12} 
\end{equation}
Where $f(\lambda _k)$ is the $k^{th}$ eigenvalue of Laplacian matrix. \\

Definition 3: In the second-order consensus problem, each node has two states. Network coherence of a second order system is measured by
\begin{equation}
{H^{(2)}  = \frac{1}{{2N}}\sum\limits_{k = 2}^N {\frac{1}{{f(\lambda _k)^2 }}} } \label{e13}
\end{equation}

Definition 4: Maximum Communication Time-Delay \cite{qi2018consensus}, \cite{olfati2004consensus} measures the ability of consensus algorithm resilient to maximum communication delay between nodes and it is expressed as
\begin{equation}
T_{\max }  = \frac{\pi }{{2f(\lambda _{n-1}(L) })} \label{e14}
\end{equation}
Where $f(\lambda _{n-1})$ is the largest eigenvalue of Laplacian matrix. \\

In this work, we use the \textit{best constant weights} algorithm to derive the closed-form expressions of convergence time as this algorithm gives the fastest convergence rate among the other uniform weight methods \cite{toulouse2015}, \cite{2011local}. \\

\begin{algorithm}
Derive the expression for eigenvalues of Laplacian matrix. \\ 
Compute the second smallest eigenvalue of Laplacian matrix ($f(\lambda _1 \left( L \right))$) and largest eigenvalue of Laplacian matrix ($f(\lambda _{n - 1} \left( L \right))$). \\
Obtain `\textit{h}' using 
\begin{equation}
{\left| {1 - hf(\lambda _1 \left( L \right)} \right)| = \left| {1 - hf(\lambda _{n - 1} \left( L \right))} \right|}
\label{8}
\end{equation}\\
Substitute the `\textit{h}'  in $\left| {1 - hf(\lambda _1 \left( L \right)} \right)|$ to obtain the convergence parameter ($\gamma$). \\
Finally, convergence time ($T$) can be calculated by 
\begin{equation}
T = \frac{1}{{\ln \left( {\frac{1}{\gamma }} \right)}}.
\label{9}
\end{equation}
\caption{{\sc}Best Constant  Weights Algorithm}
\label{algo:max}
\end{algorithm}

\section{q-triangular r-regular ring networks and related matrices}\label{section1}
\begin{figure}[tbp]
\centering
\includegraphics[totalheight=5cm]{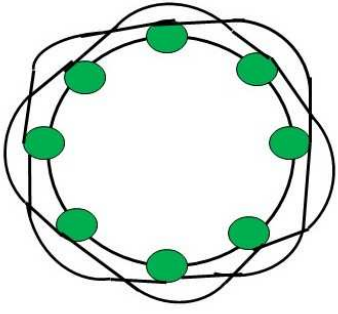}
\caption{4-regular ring network}
\label{fig:1}
\end{figure}

\begin{figure}[tbp]
\centering
\includegraphics[totalheight=5.1cm]{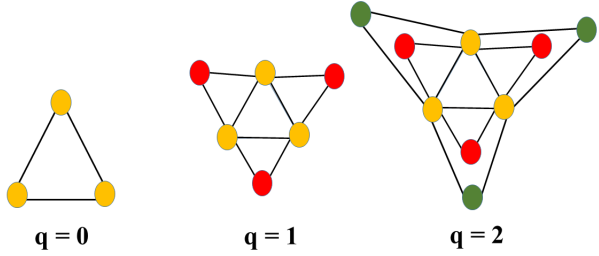}
\caption{q-triangular 2-regular ring network}
\label{fig:2}
\end{figure}

In this section, we derive the eigenvalues of Laplacian matrix for q-triangular r-regular ring networks. We can see r-regular ring network in Fig. 1 for $r=4$ and q-triangular r-regular ring network in Fig. 2 for r=2. \\

Lemma 1\emph{\cite{bapat2010graphs}} : Let $G$ be a simple connected graph with $n$ vertices. Then the rank of its incidence matrix $B$ is $n - 1$ if $G$ is bipartite and $n$ otherwise. \\

Lemma 2\label{negative}\emph{\cite{bapat2010graphs}} : Let $G$ be a bipartite graph with adjacency matrix $A$. If $\lambda$ is an eigenvalue of $A$ with multiplicity $k$, then $-\lambda$ is also an eigenvalue
of $A$ with multiplicity $k$. \\

Let the graph $G$ be bipartite and $r$-regular and $V(G)=V_1 \cup
V_2$ be a bipartition of the vertex set. Then from the proof of
Lemma \ref{negative} in \cite{bapat2010graphs} we get,
\begin{eqnarray*} A(G)\begin{bmatrix}  J_1\\
-J_2\end{bmatrix}=-r\begin{bmatrix}  J_1\\
-J_2\end{bmatrix},
\end{eqnarray*} where $J_1$ and $J_2$ are all one vectors of orders $|V_1|$ and $|V_2|$ respectively. Now as $L(G)=rI_n-A(G)$ so
\begin{eqnarray} \label{larl} L(G)\begin{bmatrix}  J_1\\
-J_2\end{bmatrix}=2r\begin{bmatrix}  J_1\\
-J_2\end{bmatrix}
\end{eqnarray} and $2r$ is the largest eigenvalue of $L(G)$ corresponding to the eigenvector $\begin{bmatrix}  J_1\\
-J_2\end{bmatrix}$.
We note that in each row of the matrix $B^T$,
there are two 1s. One is corresponding to a vertex in $V_1$ and
another is  corresponding to a vertex in $V_2$. \\
Therefore, we have
\begin{eqnarray} \label{beqn} B^T\begin{bmatrix}  J_1\\
-J_2\end{bmatrix}=\begin{bmatrix}  0\\
0\end{bmatrix}.
\end{eqnarray}
\begin{theorem}
Let $G$ be a simple, connected, $r$-regular graph with $n$ nodes and $m$ edges. Let $0=\lambda_1\leq \lambda_2\cdots\leq \lambda_n$ be the eigenvalues of $L(G)$ and $x_1,x_2,\cdots,x_n$ be their corresponding eigenvector.
Let $\mu_k=(qr+\lambda_k-2)^2+4q(2r-\lambda_k)$ and $y=\frac{qr+\lambda_k-2\mp \sqrt{\mu_k}}{2q(2r-\lambda_k)}$. Then\\
if $G$ is non-bipartite: $\frac{1}{2}[qr+\lambda_k+2\pm \sqrt{\mu_k}]$, are eigenvalues of $L(R_q(G))$ corresponding to the eigenvector $\begin{bmatrix}  x_i\\
yB^Tx_i\\
\vdots \\
yB^Tx_i\\\end{bmatrix}$ for $i=1,2,\cdots,n$
 and 2 is an eigenvalue of $L(R_q(G))$ with multiplicity $mq-n$, corresponding to the eigenvectors $\begin{bmatrix}  0\\
V_j\end{bmatrix}$ for $j=1,2,\cdots,mq-n$, where $\{V_1,V_2,\cdots,V_{mq-n}\}$ is a basis of the null space of the matrix $C=(B~ B ~\cdots ~B)$.\\
if $G$ is bipartite: $\frac{1}{2}[qr+\lambda_k+2\pm \sqrt{\mu_k}]$, are eigenvalues of $L(R_q(G))$ corresponding to the eigenvector $\begin{bmatrix}  x_i\\
yB^Tx_i\\
\vdots \\
yB^Tx_i\\\end{bmatrix}$ for $i=1,2,\cdots,n-1$,  2 is an eigenvalue of $L(R_q(G))$ with multiplicity $mq-n+1$, corresponding to the eigenvectors $\begin{bmatrix}  0\\
U_j\end{bmatrix}$ for $j=1,2,\cdots,mq-n+1$, where
$\{U_1,U_2,\cdots,U_{mq-n+1}\}$ is a basis of the null space of the
matrix $C=(B~ B ~\cdots ~B)$ and $r(q+2)$ is an eigenvalue of
$L(R_q(G))$ corresponding to the eigenvector $\begin{bmatrix}  x_n\\
0\end{bmatrix}$.
\end{theorem}
\begin{proof} Assume that the construction of $R_q(G)$ from $G$ happens by $q$ consecutive steps.
First step we construct $R_1(G)$ by adding $m$ number vertices in
$G$, then in the $i$th step we construct $R_i(G)$ by adding $m$
number vertices in $R_{i-1}(G)$ for $i=2,3,\cdots,q$. Let $V^{(i)}$
is the set of newly added vertices in $i$th step. Then we have
$\displaystyle V(R_q(G))=V(G)\cup V^{(1)}\cup V^{(2)}\cdots\cup
V^{(q)}$. With this ordering of vertices in $R_q(G)$ we get $$L(R_q(G))=\begin{bmatrix}  qD(G)+L(G) & -B & -B &\cdots & -B\\
-B^T & 2I_m & 0 & \cdots & 0\\
-B^T & 0 & 2I_m & \cdots & 0\\
\vdots &\vdots &\vdots &\ddots &\vdots \\
-B^T & 0 & 0 & \cdots & 2I_m\\\end{bmatrix}.$$
As $G$ is $r$-regular, $D(G)=rI_n$. Let $f(\lambda_k)=\frac{1}{2}[qr+\lambda_k+2\pm \sqrt{\mu_k}]$.\\
When $G$ is non-bipartite, we have for each $i\in \{1,2,\cdots,n\},$
\begin{eqnarray}\nonumber \label{eq1} L(R_q(G))\begin{bmatrix}  x_i\\
yB^Tx_i\\
yB^Tx_i\\
\vdots \\
yB^Tx_i\\\end{bmatrix}\nonumber&=&\begin{bmatrix}  q r x_i+\lambda_i x_i-q y BB^T x_i\\
-B^Tx_i+2yB^Tx_i\\
-B^Tx_i+2yB^Tx_i\\
\vdots \\
-B^Tx_i+2yB^Tx_i\\\end{bmatrix} \\ \nonumber&=&\begin{bmatrix}  (q r+\lambda_k-2q yr+qy\lambda_k) x_i\\
(2y-1)B^Tx_i\\
(2y-1)B^Tx_i\\
\vdots \\
(2y-1)B^Tx_i\\\end{bmatrix}\\
\nonumber &=& \begin{bmatrix} \frac{1}{2}[qr+\lambda_k+2\pm \sqrt{\mu_k}] x_i\\
\frac{1}{2}[qr+\lambda_k+2\pm \sqrt{\mu_k}]yB^Tx_i\\
\frac{1}{2}[qr+\lambda_k+2\pm \sqrt{\mu_k}]yB^Tx_i\\
\vdots \\
\frac{1}{2}[qr+\lambda_k+2\pm \sqrt{\mu_k}]yB^Tx_i\\\end{bmatrix}\\ &=&f(\lambda_k)\begin{bmatrix}  x_i\\
yB^Tx_i\\
yB^Tx_i\\
\vdots \\
yB^Tx_i\\\end{bmatrix}.
\end{eqnarray}
Thus we get for $i=1,2,\cdots,n$, $f(\lambda_k)$ are the eigenvalues
of $L(R_q(G))$. Now as the graph $G$ is bipartite, so $rank(B)=n$,
then the matrix $C=(B~ B~ \cdots~ B)$ which is of order $n\times
mq$, is also have $rank~n$. Therefore $dim(Ker(C))=mq-n$. Let
$V_1,V_2,\cdots V_{mq-n}$ be a basis of $Ker(C)$. Then for each
$j\in \{1,2,\cdots,mq-n\},$ we have
\begin{eqnarray} \label{eq2} L(R_q(G))\begin{bmatrix}  0\\
V_j\end{bmatrix}=2\begin{bmatrix}  0\\
V_j\end{bmatrix}.
\end{eqnarray}
Thus we get 2 is an eigenvalue of $L(R_q(G))$ with multiplicity $mq-n$.\\
When $G$ is bipartite, we have $\lambda _n=2r$. So the term $y$ in equation (\ref{eq1}) is not defined for $\lambda _n=2r$. But for all $i\in \{1,2,\cdots,n-1\},$ equation (\ref{eq1}) is satisfied. Thus $f(\lambda_k)$ are the eigenvalues of $L(R_q(G))$ for each $i\in \{1,2,\cdots,n-1\}$. Also, as $G$ is bipartite, so $rank(B)=n-1$, therefore $dim(Ker(C))=mq-n+1$. So if $U_1,U_2,\cdots U_{mq-n+1}$ is a basis of $Ker(C)$, then for each $j\in \{1,2,\cdots,mq-n+1\},$ we have
\begin{eqnarray*} L(R_q(G))\begin{bmatrix}  0\\
U_j\end{bmatrix}=2\begin{bmatrix}  0\\
U_j\end{bmatrix}.
\end{eqnarray*}
Thus we get 2 is an eigenvalue of $L(R_q(G))$ with multiplicity $mq-n+1$. Also we have
\begin{eqnarray*} L(R_q(G))\begin{bmatrix}  x_n\\
0\\
0\\
\vdots \\
0\\\end{bmatrix}=\begin{bmatrix}  (q r+2r)x_n\\
-B^Tx_n\\
-B^Tx_n\\
\vdots \\
-B^Tx_n\\\end{bmatrix}=(q r+2r)\begin{bmatrix}  x_n\\
0\\
0\\
\vdots \\
0\\\end{bmatrix}.
\end{eqnarray*}
Therefore $q r+2r$ is an eigenvalue of $L(R_q(G))$.
\end{proof}
\section{convergence analysis for q-triangular r-regular ring network}
In this section, we derive the explicit expressions of Convergence Time, First Order Network Coherence, Second Order Network Coherence, and Maximum Communication Time-Delay. 
\begin{theorem}
Convergence Time of q-triangular r-regular ring network for average consensus algorithm is 
\begin{equation}
T=\frac{1}{{\ln \left( {\frac{{(q + 1)r + 3 - l_1 }}{{\sqrt {\left( {1 - (q + 1)r + l_1 } \right)^2  + 4q(r - 1 + l_1 )} }}} \right)}}, 
\end{equation}
where \begin{equation} \nonumber
l_1  = \frac{{\sin \frac{\pi }{n}(r + 1)}}{{\sin \frac{\pi }{n}}}.
\end{equation}
\end{theorem}
\begin{proof}
Eigenvalues of Laplacian matrix for $r$-regular ring network is 
\begin{equation}
\lambda _k  = r - 2\sum\limits_{j = 1}^{\frac{r}{2}} {\cos \frac{{2\pi kj}}{n}} 
\end{equation}\label{20}
From Dirchilet Identity $
1 + 2\sum\limits_{j = 1}^r {\cos \left( {jx} \right) = \frac{{\sin \left( {r + \frac{1}{2}} \right)x}}{{\sin \frac{x}{2}}}}$, we can further simplify the above expression as
\begin{equation}
\lambda _k  =r + 1 - \frac{{\sin \left( {r + \frac{1}{2}} \right)x}}{{\sin \frac{x}{2}}} \label{e1}
\end{equation}
where $n$ is number of nodes and $r$ is node degree. \\
From Theorem 1, we can write the eigenvalues of Laplacian matrix for q-triangular regular network can be written as
\begin{equation}
 f(\lambda _k )= \frac{1}{2}\left( {qr + \lambda _k  + 2 \pm \sqrt {\left( {qr + \lambda _k  - 2} \right)^2  + 4q\left( {2r - \lambda _k } \right)} } \right) \label{e2}
\end{equation}
We substitute the (\ref{e1}) in (\ref{e2}) to obtain the eigenvalues of Laplacian matrix for q-triangular r-regular ring network as 
\begin{equation}
f(\lambda _k ) = \frac{1}{2}\left( {q_2  + 3 - l \pm \sqrt {\left( {q_2  - 1 - l} \right)^2  + 4q(r - 1 + l)} } \right)
\end{equation},
where $l_2  = \frac{{\sin \frac{{(r + 1)\pi k}}{n}}}{{\sin \frac{{\pi k}}{n}}}$ and $(q + 1)r = q_2 $. \\
In the $\lambda _0 ,\lambda _1 ,...\lambda _{n - 1} $, we have observed $\lambda _1$ is the second smallest eigenvalue and $\lambda _{n - 1}$ is the largest eigenvalue of Laplacian matrix. \\
Hence, we can write the largest eigenvalue of Laplacian matrix of q-triangular r-regular ring network as 
\begin{equation}
f(\lambda _{n - 1} ) = \frac{1}{2}\left( {q_2  + 3 - l_3  + \sqrt {\left( {q_2  - 1 - l_3 } \right)^2  + 4q(r - 1 + l_3 )} } \right) \label{e4}
\end{equation},
where $l_3  = \frac{{\sin \frac{{(r + 1)\pi }}{n}}}{{\sin \frac{\pi }{n}}}$ \\
Similarly, we can write the smallest eigenvalue of Laplacian matrix of q-triangular r-regular ring network as 
\begin{equation}
f(\lambda _1 ) = \frac{1}{2}\left( {q_2  + 3 - l_3  - \sqrt {\left( {q_2  - 1 - l_3 } \right)^2  + 4q(r - 1 + l_3 )} } \right)
\end{equation}
From best constant algorithm, we can obtain the convergence parameter as
\begin{equation}
\gamma  = \frac{{\sqrt {\left( {(q + 1)r - 1 - \frac{{\sin \frac{{(r + 1)\pi }}{n}}}{{\sin \frac{\pi }{n}}}} \right)^2  + 4q\left( {r - 1 + \frac{{\sin \frac{{(r + 1)\pi }}{n}}}{{\sin \frac{\pi }{n}}}} \right)} }}{{(q + 1)r + 3 - \frac{{\sin \frac{{(r + 1)\pi }}{n}}}{{\sin \frac{\pi }{n}}}}} \label{e3}
\end{equation}
Substituting the $\gamma$ value in (\ref{e11}) proves the theorem.
\end{proof}
\begin{theorem}
Network Coherence of q-triangular r-regular ring network for first order system is 
\begin{equation}
H^{(1)}  = \frac{1}{{2N}}\sum\limits_{k = 1}^{(N-1)} {\frac{2}{{(q + 1)r + 3 - l + c_1 }}},
\end{equation}
and
second order system is
\begin{equation}
H^{(2)}  = \frac{1}{{2N}}\sum\limits_{k = 1}^{(N-1)} {\frac{2}{{\left( {(q + 1)r + 3 - l + c_1 } \right)^2 }}},
\end{equation},
where $c_1=\sqrt {\left( {1 - (q + 1)r + l} \right)^2  + 4q(r + l - 1)}$, $l = \frac{{\sin \frac{\pi }{n}(r + 1)k}}{{\sin \frac{\pi }{n}}}$. 
\end{theorem}
\begin{proof}
Substitute the (\ref{e2}) into (\ref{e12}) and (\ref{e13}) to prove the Theorem. 
\end{proof}

\begin{theorem}
Maximum Communication Time-Delay of q-triangular r-regular ring network for average consensus algorithm is 
\begin{equation}
R_{\max }  = \frac{\pi }{{(q + 1)r + 3 - l - c_1 }},
\end{equation}
$c_1=\sqrt {\left( {1 - (q + 1)r + l} \right)^2  + 4q(r + l - 1)}$, $l = \frac{{\sin \frac{\pi }{n}(r + 1)k}}{{\sin \frac{\pi }{n}}}$.
\end{theorem}
\begin{proof}
Substitute the (\ref{e4}) into (\ref{e14}) proves the theorem.
\end{proof}
\section{Results and Discussion}
In this section, we present the numerical results to investigate the effect of node degree,  network size, and triangulation parameter on convergence time, first order network coherence, second order network coherence, and maximum communication time-delay of average consensus algorithm. In Fig. 3, we plot the convergence time against triangulation parameter for $n$=100. We have observed the convergence time linearly increases with the triangulation parameter $q$ and convergence time reduces with the increase in $r$ values. Increase in degree will leads to more participation among IoT nodes. This will naturally reduces the convergence time. To observe the effect of network size on convergence time, we plot the Fig. 4 for $r$=4. It is noted that convergence time is drastically increasing with the network size. For $r$=50, we plot the maximum communication time-delay against triangulation parameter $q$ for different network sizes in Fig. 5, and observed that maximum communication time-delay exponentially reduces with the $q$. This is because of nodes are able to reach the long distance nodes due to triangulation operation. To study the effect of number of node degree on maximum communication time-delay, we plot Fig. 6 for $n$=100. It is observed that maximum communication time-delay decreases with $r$ values. In Fig. 7, we plot the first order network coherence versus triangulation parameter for $r=4$ and observed that network coherence exponentially decreases with the triangulation parameter and decreases with the node degree. To observe the effect of node degree on second order network coherence, we plot the Fig. 8. We have observed that there is a very little increase in second order network coherence with the increase in $r$ values. In Fig. 9, we plot the first order consensus versus $q$ for $r$=5 and observed that first order network coherence increases with the network size. As shown in Fig. 10, the effect of network size on second order network coherence is drastically reduced. 
\begin{figure}
\centering
\includegraphics[width=9cm,height=7cm]{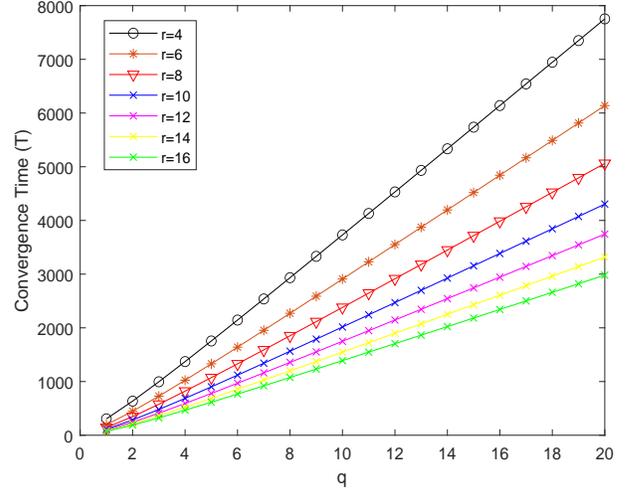}
\caption{Convergence Time versus $q$ for Average Gossip Algorithm ($n$=100).}
\label{fig:3}
\end{figure}
\begin{figure}
\centering
\includegraphics[width=9cm,height=7cm]{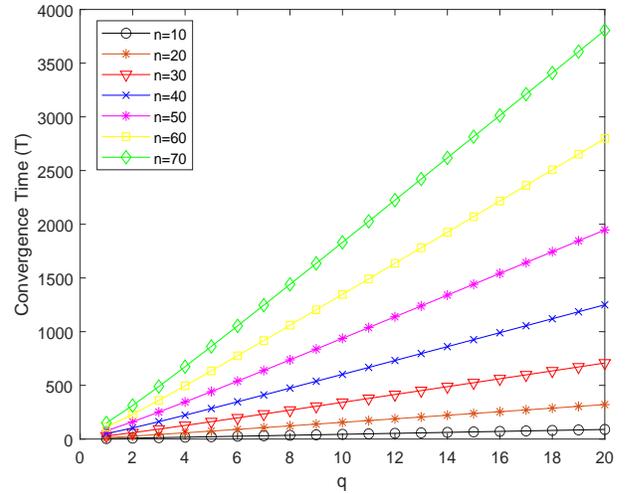}
\caption{Convergence Time versus $q$ for Average Gossip Algorithm ($r$=4).}
\label{fig:4}
\end{figure}
\begin{figure}
\centering
\includegraphics[width=9cm,height=7cm]{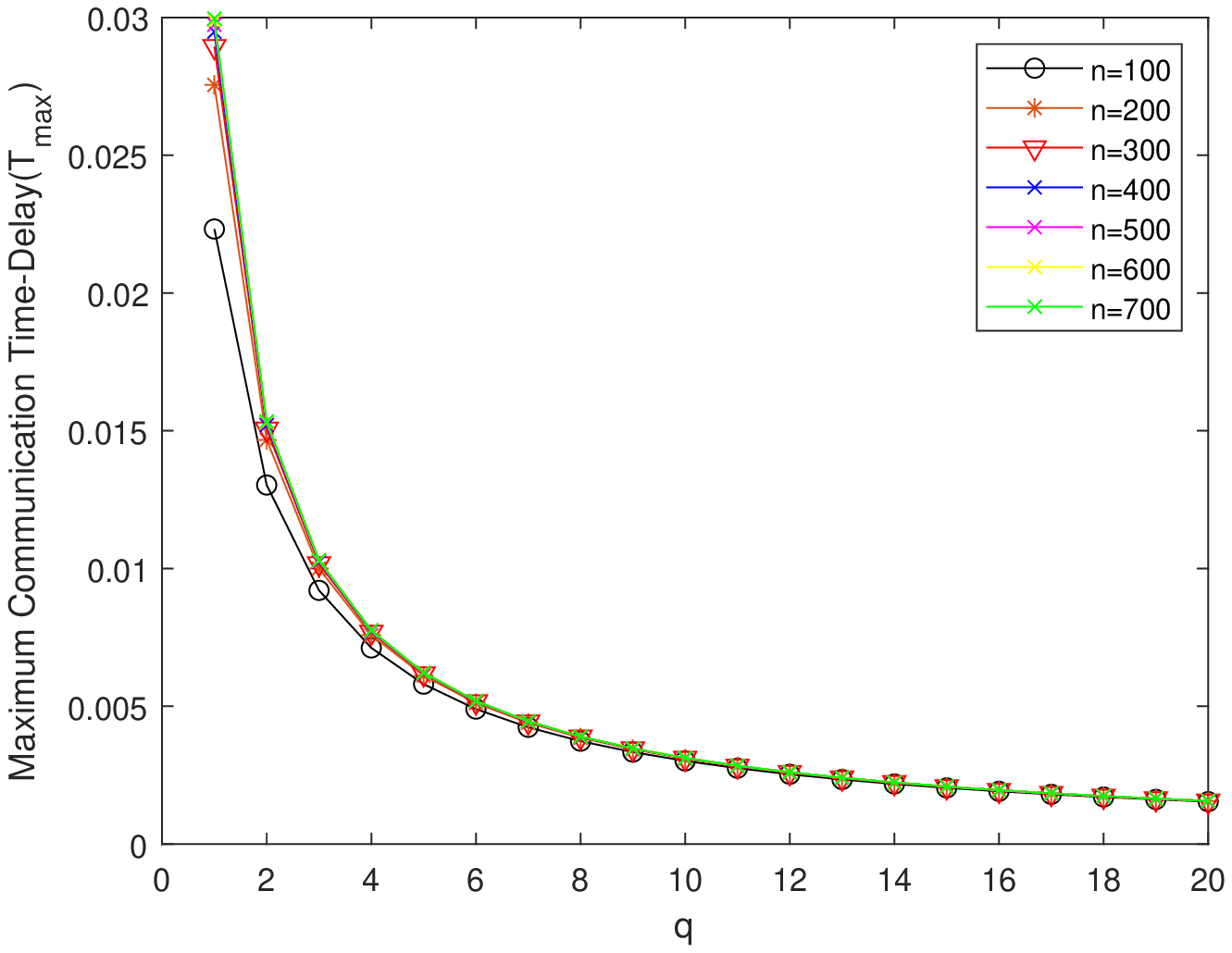}
\caption{Maximum Communication Time-Delay versus $q$ for Average Gossip Algorithm ($r$=50).}
\label{fig:5}
\end{figure}
\begin{figure}
\centering
\includegraphics[width=9cm,height=7cm]{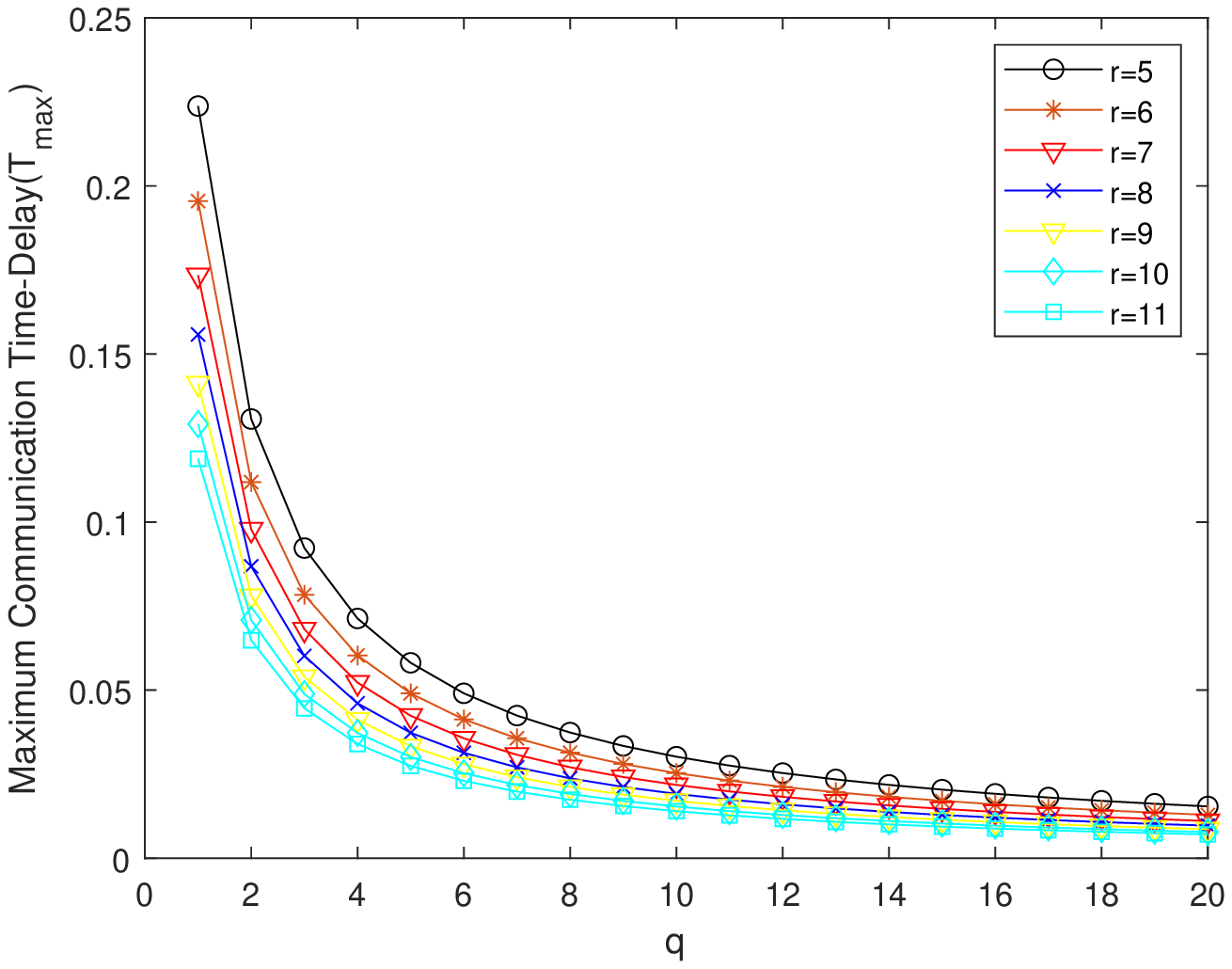}
\caption{Maximum Communication Time-Delay versus $q$ for Average Gossip Algorithm ($n$=100).}
\label{fig:6}
\end{figure}
\begin{figure}
\centering
\includegraphics[width=9cm,height=7cm]{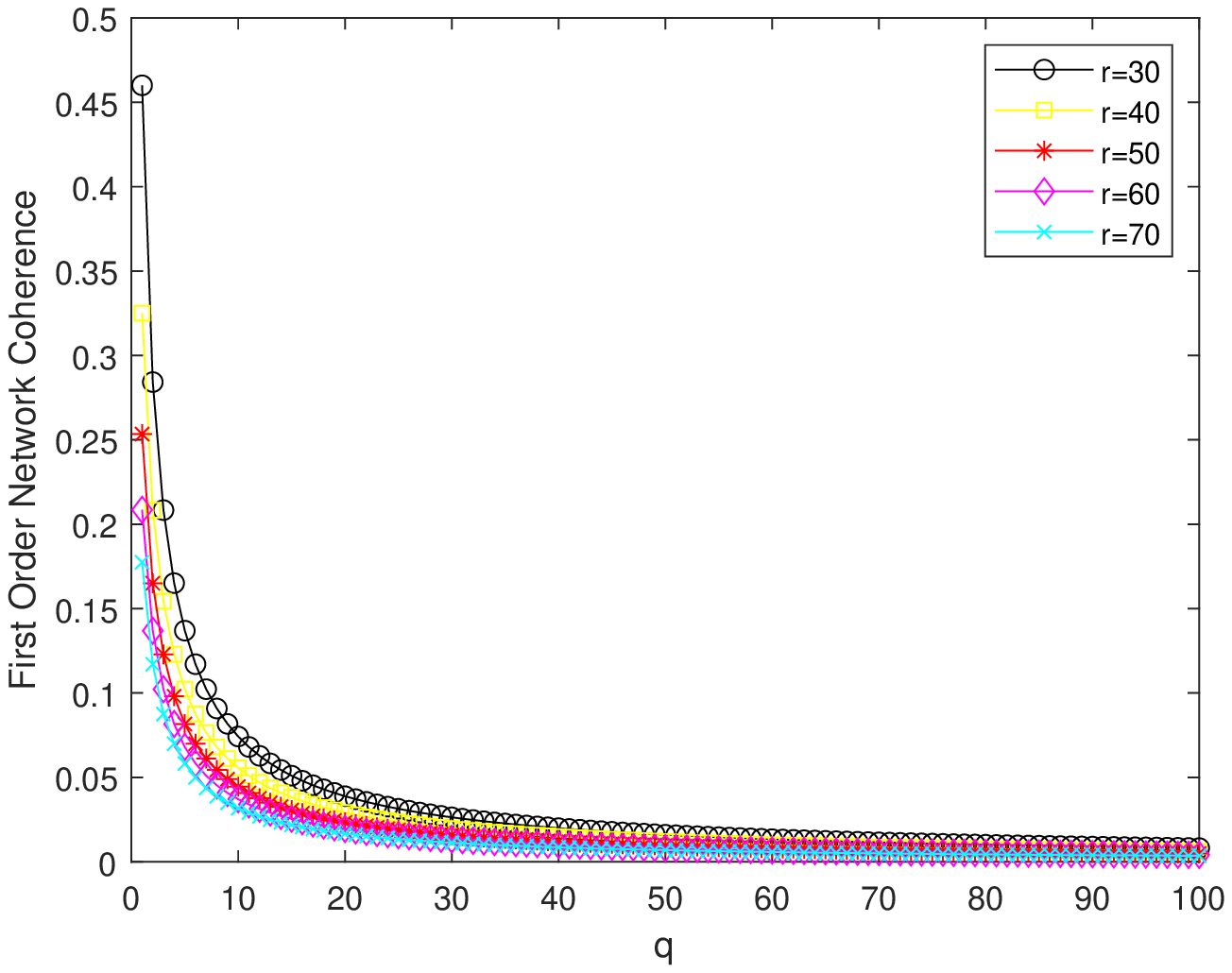}
\caption{First Order Coherence versus $q$ for Average Gossip Algorithm ($r$=4).}
\label{fig:7}
\end{figure}
\begin{figure}
\centering
\includegraphics[width=9cm,height=7cm]{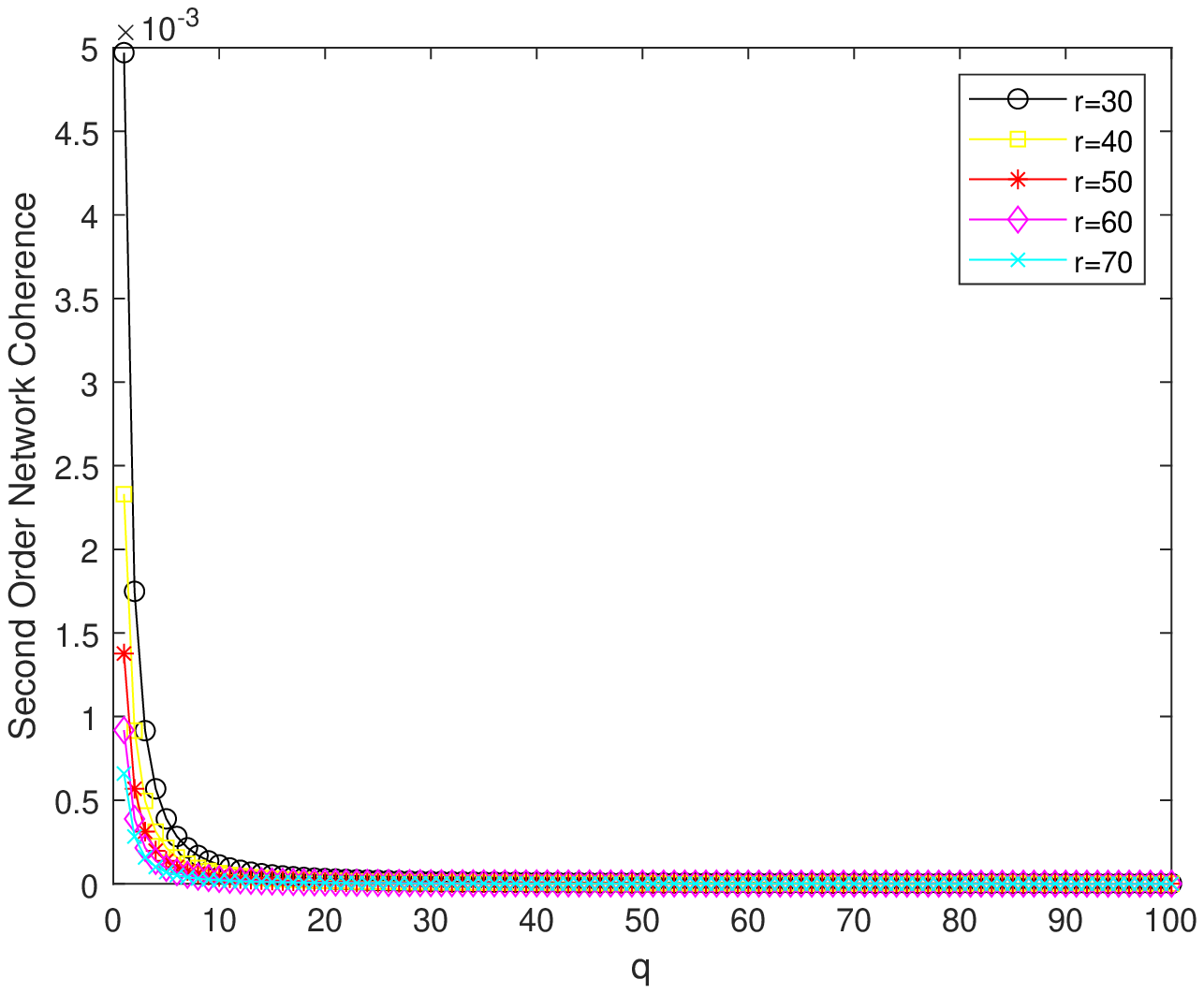}
\caption{Second Order Coherence versus $q$ for Average Gossip Algorithm ($n$=100).}
\label{fig:8}
\end{figure}
\begin{figure}
\centering
\includegraphics[width=9cm,height=7cm]{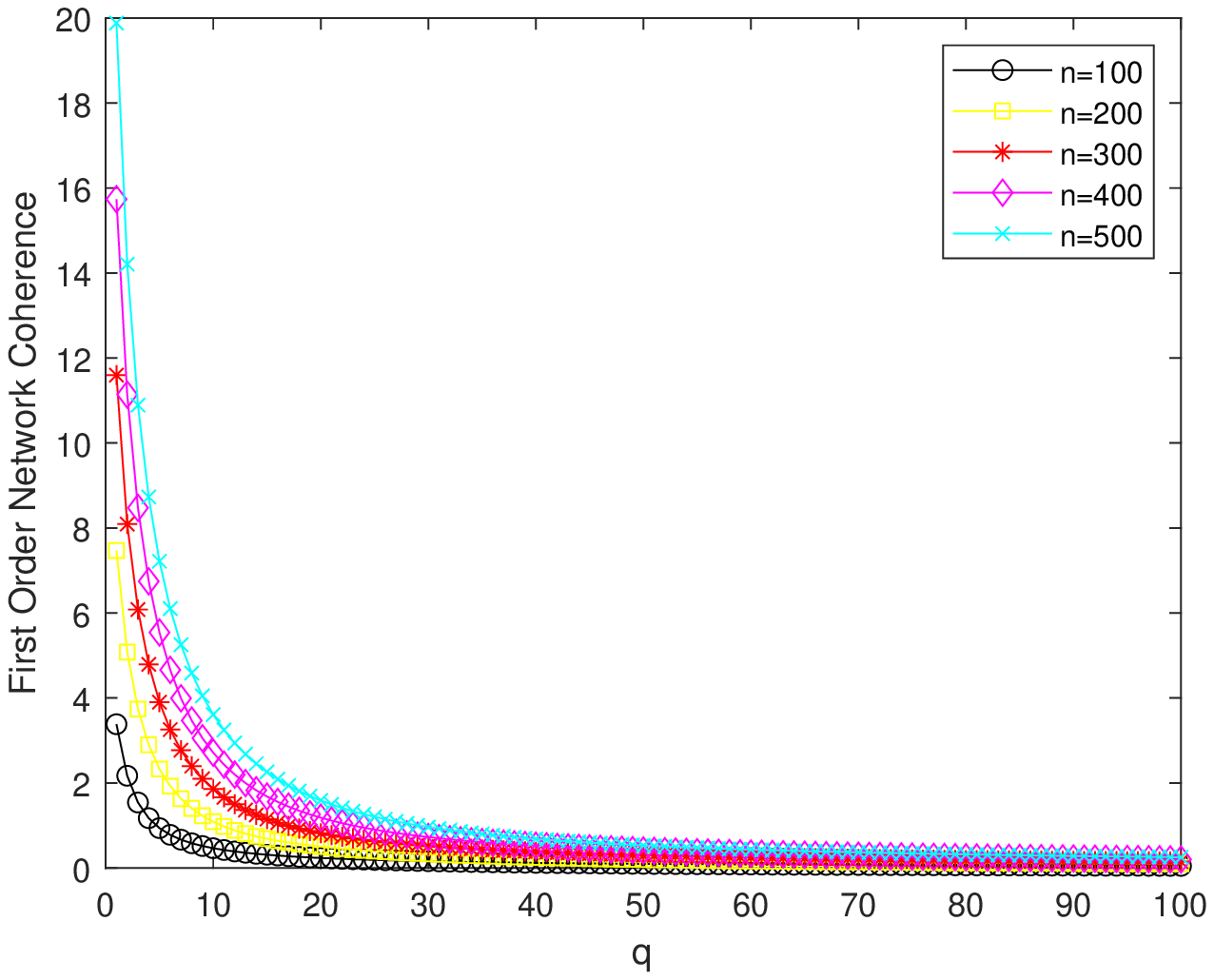}
\caption{First Order Coherence versus $q$ for Average Gossip Algorithm ($r$=5).}
\label{fig:9}
\end{figure}
\begin{figure}[H]
\centering
\includegraphics[width=9cm,height=7cm]{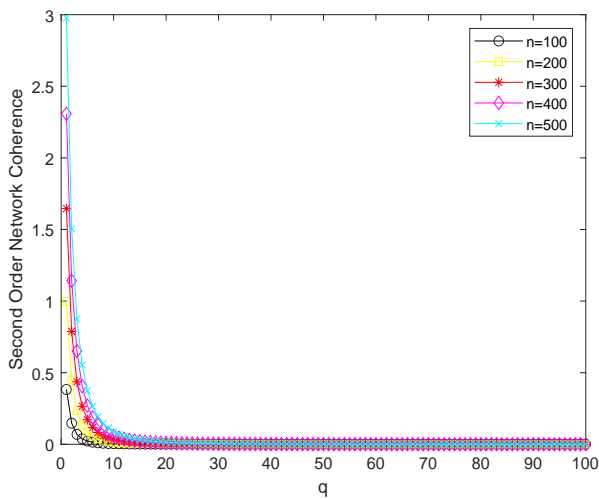}
\caption{Second Order Coherence versus $q$ for Average Gossip Algorithm ($r$=5).}
\label{fig:10}
\end{figure}
\section{Conclusions}
In this paper, we have modeled the IoT network as an q-triangular r-regular network and studied the consensus algorithms, with an emphasis on the convergence time, network coherence, and maximum communication time-delay. We first provided the eigenvalues of Laplacian matrix for q-triangular r-regular networks. We then derived the explicit expressions of convergence time, network coherence, and maximum communication time-delay for q-triangular r-regular ring networks. We studied numerically the convergence time, first order coherence, second-order coherence, and maximum communication time-delay with respect to network size, triangulation parameter, and node degree. Our results indicate that proposed network topology is resistant to noise and communication time-delay of average consensus algorithms for large-scale IoT networks. We argued that the scale-free and small-world structure of the q-triangulation networks is responsible for the robustness. Future work should include the unveiling the effects of energy consumption and position of nodes on the performance metrics of consensus algorithms for robust IoT networks.



%
\section*{Acknowledgement}
Fouzul Atik would like to thank the Department of Science and Technology, Government of India, for the financial support (Start-up Research Grant (SRG/2019/000839)).

%
\bibliographystyle{IEEEtran}
\bibliography{References}

\end{document}